\documentclass[conference]{IEEEtran}

\IEEEoverridecommandlockouts
\usepackage{cite}
\usepackage{amsmath,amssymb,amsfonts}
\usepackage{algorithmic}
\usepackage{graphicx}
\usepackage{textcomp}
\usepackage{xcolor}
\usepackage{booktabs,tabularx,makecell,array}
\renewcommand\theadfont{\bfseries}

\def\BibTeX{{\rm B\kern-.05em{\sc i\kern-.025em b}\kern-.08em
    T\kern-.1667em\lower.7ex\hbox{E}\kern-.125emX}}
\begin{document}

\title{Power Delivery for Cryogenic Scalable Quantum Applications: Challenges and Opportunities}
{\footnotesize}

\author{
Yating Zou$^{*}$, Batuhan Keskin$^{*}$, Gregor G. Taylor$^{*}$, Zenghui Li$^{\ddagger}$, \\
 Jie Wang$^{\ddagger}$, Eduard Alarcón$^{\dagger}$, Fabio Sebastiano$^{\ddagger}$, Masoud Babaie$^{\ddagger}$, Edoardo Charbon$^{*}$\\[3pt]
$^{*}$Department of Microsystems and Microelectronics, École Polytechnique Fédérale de Lausanne, Switzerland\\
$^{\dagger}$School of Telecommunications, Technical University of Catalunya, Spain\\
$^{\ddagger}$Department of Quantum and Computing Engineering \& QuTech, Delft University of Technology, The Netherlands
}

\maketitle

\begin{abstract}
Quantum technologies offer unprecedented capabilities in computation and secure information transfer. Their implementation requires qubits to operate at cryogenic temperatures (CT) while control and readout electronics typically still remains at room temperature (RT). As systems scale to millions of qubits, the electronics should also operate at CT to avoid a wiring bottleneck. However, wired power transfer from RT for such electronics introduces severe challenges, including thermal load between cooling stages, Joule heating, noise coupling, and wiring scalability.  This paper addresses those challenges by evaluating several candidate architectures for scalable power transfer in the dilution frige: high-voltage (HV) wired power transfer, radiative wireless transfer, non-radiative wireless transfer, and a hybrid HV and non-radiative transfer. These architectures are analyzed in terms of thermal load, power loss, heating, coupling noise, power density, scalability, reliability, and complexity. Comparative analysis demonstrates the trade-offs among these architectures, while highlighting HV non-radiative transfer as a promising candidate for scalable quantum systems.
\end{abstract}

\begin{IEEEkeywords}
Quantum applications, power, cryogenic, Joule heating.
\end{IEEEkeywords}

\section{Introduction}
\begin{figure}
\centering
\includegraphics[width=\columnwidth]{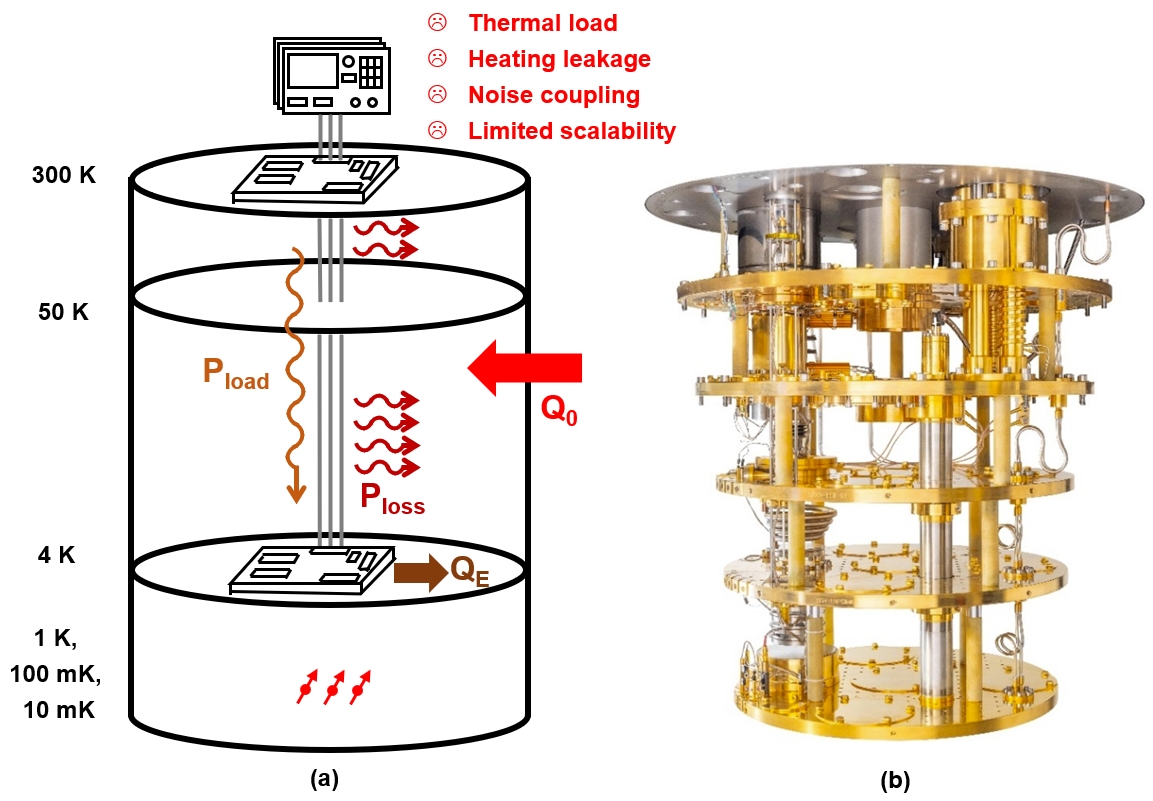}
\caption{Existing wired power transfer in dilution fridge: (a) diagram of wired power transfer (b) prototype of dilution fridge \cite{zu2022dilution}.}
\label{Power_transfer_in_DF}
\end{figure}

Quantum applications such as computing and communication promise unprecedented computational and security advantages. Quantum devices and qubits operate at cryogenic temperatures (CT), while control and readout electronics typically operate at room temperature (RT) outside the dilution refrigerator or cryostat. As quantum computing scales up, this temperature disparity introduces significant challenges, particularly in data transfer between RT and CT. One solution has been proposed to shift electronics from RT to CT to enable efficient qubit operation \cite{charbon2016,controller2020,readout2024}. However, the existing wired power transfer from RT for electronics at CT shown in Fig. \ref{Power_transfer_in_DF} remains a major challenge when scaling quantum computers to large numbers of qubits due to (i) thermal load arising from temperature gradients, (ii) power loss and heating by Joule effect, (iii) noise coupling from RT to CT, and (iv) limited scalability caused by interconnect wiring. The scaling for power transfer is highly complex and has so far only focused primarily on individual power management circuits (e.g. cryogenic low-dropout regulators (LDOs)\cite{LDO2021}, voltage references \cite{Job2024}, current references \cite{Ray2025}). Nevertheless, the major bottleneck is expected to shift toward the architecture challenge of power transfer.

Existing wired power transfer relies on multiple twisted phosphor-bronze wires \cite{krinner2019}, which provide high thermal resistance for improved thermal isolation but suffer from high electrical resistance, or copper wires with low thermal and electrical resistance. As the quantum computer system scales to millions of qubits, the power required for controller and readout circuits increases significantly, and the thermal load from warmer to colder stages becomes non-negligible. Additionally, high electrical resistance $R_{\text{wire}}$ ($16\,\Omega @ 300\,\text{K},\ 12\,\Omega @ 4\,\text{K}$ per wire  \cite{bluefors}) of such phosphor-bronze wires causes non-negligible power loss, equivalent to $I^{2} R_{\text{wire}}$, which is dissipated as heat by Joule effect, thereby potentially destabilizing temperature control. Noise coupling from power-supply instruments at RT to CT, particularly at low frequencies, further deteriorates system performance. Finally, the impractically high density of interconnect wires for power transfer is incompatible with large-scale quantum applications.

\begin{figure}
\centering
\includegraphics[width=\columnwidth]{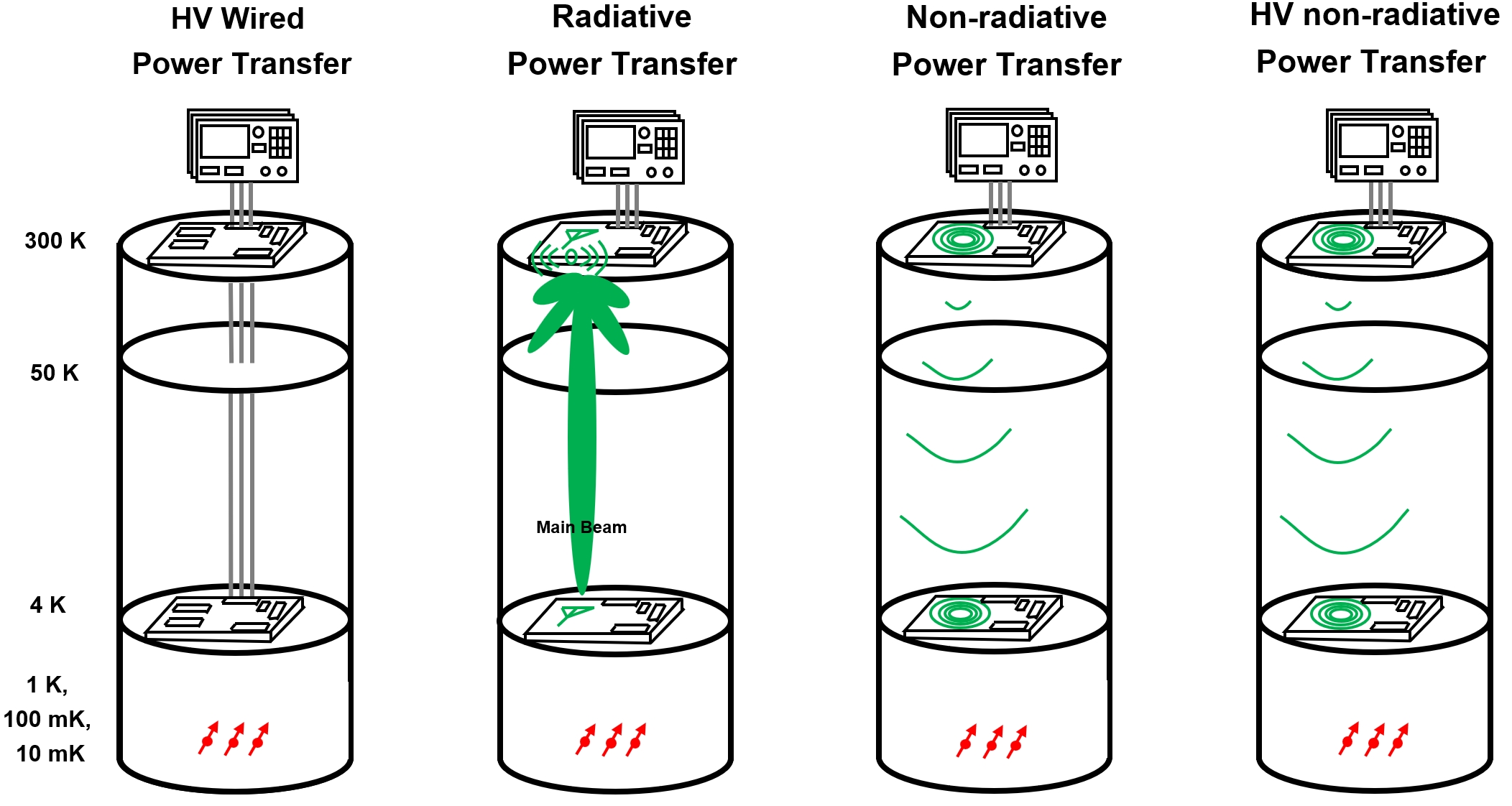}
\caption{Different power transfer architectures in and around a dilution refrigerator.}
\label{Different_ways_of_power_transfer}
\end{figure}

To realize a promising power transfer vision for future large-scale qubits or quantum devices, this paper proposes several architectures including high voltage (HV) wired power transfer, radiative power transfer, non-radiative power transfer, and a combination of HV and non-radiative power transfer, as shown in Fig. \ref{Different_ways_of_power_transfer}. These architectures are evaluated with respect to thermal load, power loss, heating, coupling noise, power density, scalability, reliability, and complexity. For each solution, the complete corresponding power management system including interconnections and electronics are also discussed.

\section{Potential Power Transfer Strategies for Scalable Quantum Applications}
The power consumption of electronics for qubits or quantum devices such as controller for superconducting qubits, spin qubits, photonic qubits, superconducting nanowire single photon detectors (SNSPDs \cite{morozovSuperconductingPhotonDetectors2021,reddySuperconductingNanowireSinglephoton2020,taylorLownoiseSinglephotonCounting2023,korzhDemonstrationSub3Ps2020}), rapid single flux quantum (RSFQ) signal processors and related logic families \cite{likharevRSFQLogicMemory1991, mukhanovScalableQuantumComputing2019} all scales with the number of qubits or devices. 

The power consumption of cryo-CMOS controllers for both spin \cite{barteeSpinqubitControlMillikelvin2025} and superconducting qubits \cite{yooDesignCharacterization4mW2023} is typically in the order of 1-10 mW per qubit. For photon detection and imaging applications, cameras with over 1000 pixels, as well as photonic quantum computing architectures containing thousands of detectors, are developed. Regarding the readout, various LNAs \cite{peng032x012Mm2Cryogenic2024, niuWidebandCryogenicReadout2022} have been demonstrated. Although low-power solutions with 20 $\mu\text{W}$ exist \cite{carboni2025}, high-performance detectors typically consume about 1-10 mW per pixel, while current supply circuits require on the order of 10 µW per channel. As researchers scale these detectors into large arrays, the bias and readout cables become increasingly unsustainable due to thermal constraints. RSFQ signal processors promise beyond-CMOS capabilities with orders of magnitude less power consumption and high clock speeds\cite{likharevRSFQLogicMemory1991}. The Josephson junction (JJ) serves as the fundamental block of these circuits, and its bias requirements necessitate substantial bias currents and corresponding cabling in large-scale implementations, resulting in siginificant heating on the cryostat \cite{dotsenkoIntegration4Stage42009}.

\begin{figure}
\centering
\includegraphics[width=\columnwidth]{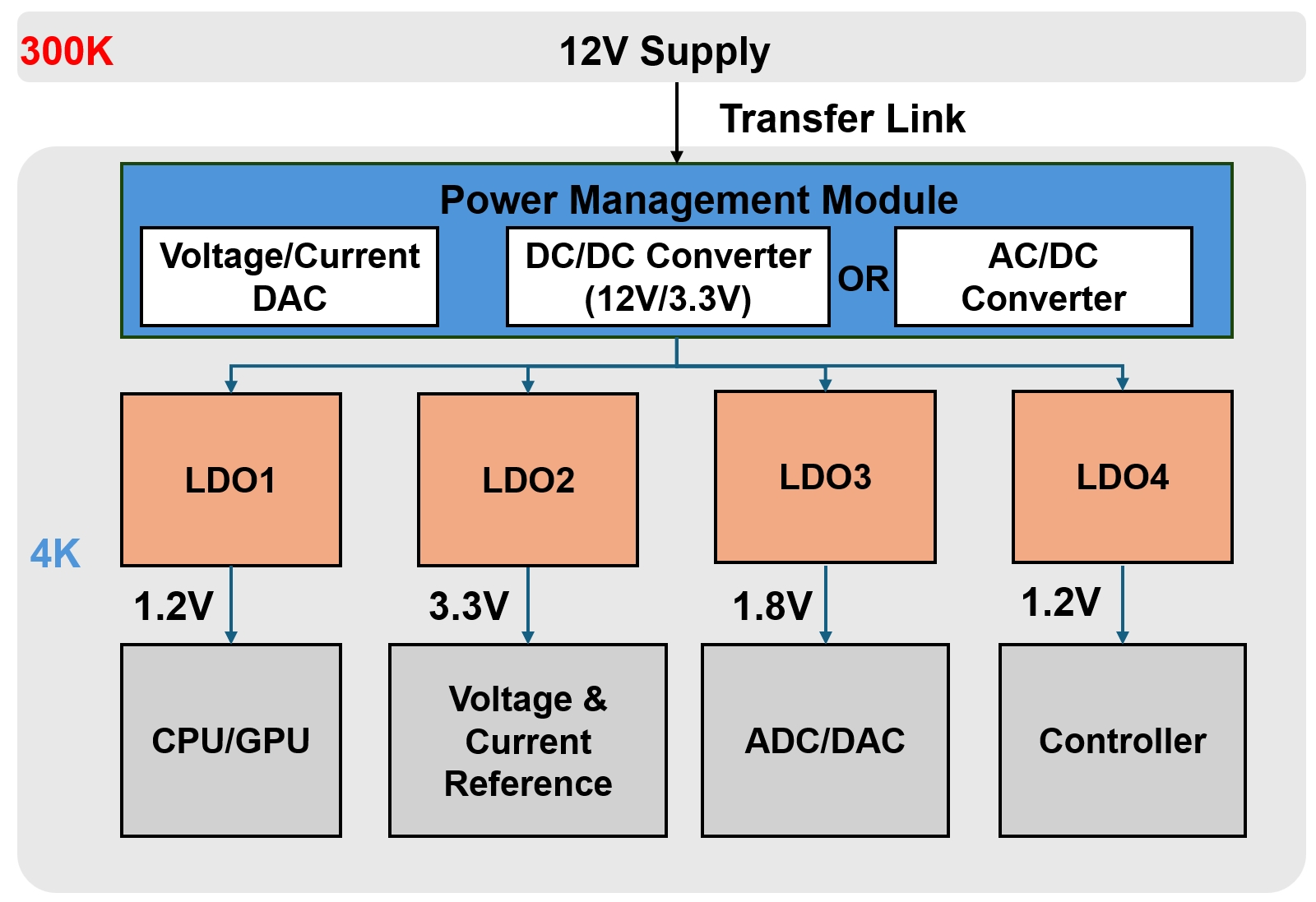}
\caption{Power management blocks.}
\label{Power_management}
\end{figure}

To power the aforementioned qubit and device electronics, four potential power transfer architectures are proposed. For large-scale qubit systems, to reduce Joule heating primarily determined by the power loss $I^{2}R_{\text{wire}}$ of the wires, a mitigation strategy is HV wired power transfer. This approach transfers high voltage from 300 K, thereby reducing the current and associated wire losses. On the receiver side at 4 K, as illustrated in Fig.~\ref{Power_management}, a cryogenic DC/DC buck converter \cite{xu2025} is required to step down the high voltage (e.g., 12 V) to lower supply levels (e.g., 3.3 V), which are subsequently distributed to various blocks such as CPU, voltage reference, ADC, and controller for qubit operation via different LDOs. However, additional switching noise from the converter, as well as noise coupling through the wires, must be carefully managed to satisfy the power supply rejection ratio (PSRR) requirements of qubit operation circuits.

Despite its advantages, HV wired power transfer still suffers from thermal load across temperature stages. A more advanced approach is to eliminate physical wiring and employ wireless power transfer. In radiative power transfer, operating above 10 GHz minimizes interference with qubits that operate in the 1–10 GHz range at millikelvin temperatures near 4 K. Transmitter antenna arrays with beamforming concentrate energy onto receiver antennas or rectennas to improve efficiency. LDOs follow the antennas and AC/DC converter, or rectennas, to provide output regulation as shown in Fig.~\ref{Power_management}. This architecture can also integrate backscattering-based data communication \cite{backscatter2025}, enabling qubit information transfer from 4 K to 300 K without a high-power local oscillator and further reducing cryogenic power consumption.

Alternatively, non-radiative power transfer also eliminates wire-related losses by operating at lower coupling frequencies ($\sim$MHz) to avoid qubit interference. A transmitter coil and capacitor form a resonant circuit that inductively couples energy to a receiver coil with a matching capacitor. The received low-frequency signal is rectified to DC with minor leakage due to lower parasitic effects. As the coils can be superconducting at 4 K, resistive heating is negligible. However, power transfer over long distances (e.g., 1 m) suffers from reduced efficiency. Several methods have been proposed to address this issue. For instance, as demonstrated in \cite{intermediatecoil2011}, the use of intermediate resonant coils with capacitors can reshape the magnetic field and extend transmission range. Accordingly, introducing an intermediate resonant stage at 50 K represents a viable strategy to further enhance efficiency without the use for active circuits.

Finally, a hybrid approach combining HV wired and non-radiative transfer offers another promising solution. In this configuration, high voltage is supplied from RT, and the resonant transmission further reduces current-related losses. This combination achieves lower transmission loss and thermal loss, making it a strong candidate for scalable quantum systems.

\section{Comparisons of Power Transfer}
To better understand the existing wired power transfer architecture and the proposed four architectures, comparisons about efficiency, heating, noise, power density, scalability, reliability, and complexity are discussed and shown in Table~\ref{tab:power_transfer}.

\renewcommand\theadfont{\bfseries}
\newcolumntype{Y}{>{\centering\arraybackslash}X}

\begin{table}[t]
\caption{Comparison of Power Transfer Methods}
\centering
\footnotesize
\setlength{\tabcolsep}{3pt}     
\renewcommand{\arraystretch}{1.25}
\begin{tabularx}{\columnwidth}{@{} c *{5}{Y} @{}}
\toprule
\makecell{\textbf{Power}\\\textbf{Transfer}\\\textbf{Architectures}} &
\thead{Wired\\Transfer} &
\thead{HV\\Transfer} &
\thead{Radiative\\Transfer} &
\makecell[c]{\textbf{Non-}\\\textbf{radiative}\\\textbf{Transfer}} &
\makecell[c]{\textbf{HV Non-}\\\textbf{radiative}\\\textbf{Transfer}} \\
\midrule
Power Loss    &  High & Very Low      & Moderate & Low & Low \\
Heating       & High& Low     & Moderate & Very Low & Very Low  \\
Noise         & Moderate & High     & Low      & Low      & Low  \\
Power Density   & Low     & Low & High      & High      & Very High  \\
Reliability & High      & Moderate      & Low     & Low     & Low \\
\bottomrule
\end{tabularx}
\label{tab:power_transfer}
\end{table}

\subsection{Power loss}
The power loss from different aspects determines the overall efficiency of power transfer system and would be dissipated as heating entering the dilution fridge.

For the wired power transfer, the dominant power loss arises from the wiring and can be expressed as
\begin{equation}
P_{\text{loss,wired}} = \left(\frac{P_{\text{RX}}}{V_{\text{RX}}}\right)^{2} R_{\text{wire}}\label{eq}
\end{equation}
where $ P_{\text{RX}}$ and $ V_{\text{RX}}$ are the required power and voltage.

For the HV wired power transfer, the power loss caused by the wiring is further reduced and can be expressed as
\begin{equation}
P_{\text{loss,HV\_wired}} = \left(\frac{P_{\text{RX}}}{V_{\text{RX\_HV}}}\right)^{2} R_{\text{wire}}\label{eq}
\end{equation}
where $ V_{\text{RX\_HV}}$ is the high received voltage.

For radiative power transfer, the stable structure of the dilution refrigerator enables precise alignment between transmitter and receiver antennas, minimizing absorption and scattering effects. The resistive loss in the receiver antennas can be neglected due to its superconducting property. The loss in the transmitter antennas $\left(\frac{P_{\text{RX}}}{V_{\text{RX}}}\right)^{2} R_{\text{TX\_ant}}$ does not contribute to heating in the fridge and is therefore not considered here. Small-aperture antennas are unsuitable for long-distance transmission due to low directivity and high free-space path loss, while large-aperture antennas mainly suffer from coupling, radiation, and resistive losses, expressed as
\begin{equation}
P_{\text{loss,rad}} = P_{\text{RX}}
\left(\frac{1}
{\eta_{\text{rad,r}} \eta_{\text{coup,ant}}}
-1\right)
  \label{eq}
\end{equation}
where $\eta_{\text{rad,r}}$ represents the radiation efficiency of antennas, and $\eta_{\text{coup,ant}}$ represents the antenna coupling efficiency. 

For the non-radiative power transfer, the free-space loss is alleviated because the waves do not undergo spherical spreading. The main losses arise from weak coupling between the coils and resistive losses in the transmitter coil. The resistive losses in the receiver and transmitter coils are similar to the case in the radiative power transfer. The total power loss can be denoted as
\begin{equation}
P_{\text{loss,non-rad}} = \frac{P_{\text{RX}}(1-\eta_{\text{coup,coil}})}{\eta_{\text{coup,coil}}} \label{eq} 
\end{equation}
where $ \eta_{\text{coup,coil}}$ is the efficiency of coupling.

For the HV non-radiative transfer, the power loss in the transmitter coil $\left(\frac{P_{\text{RX}}}{V_{\text{RX\_HV}}}\right)^{2} R_{\text{TX\_coil}}$ is reduced due to high-voltage source compared to non-radiative transfer, while the wire losses is also mitigated. As the transmitter coil loss does not cause heating at 4 K, it is neglected. Thus, the total power loss equals that of non-radiative transfer and is denoted as
\begin{equation}
P_{\text{loss,comb}} = \frac{P_{\text{RX}}(1-\eta_{\text{coup,coil}})}{\eta_{\text{coup,coil}}} \label{eq} 
\end{equation}

\begin{figure}
\centering
\includegraphics[width=\columnwidth]{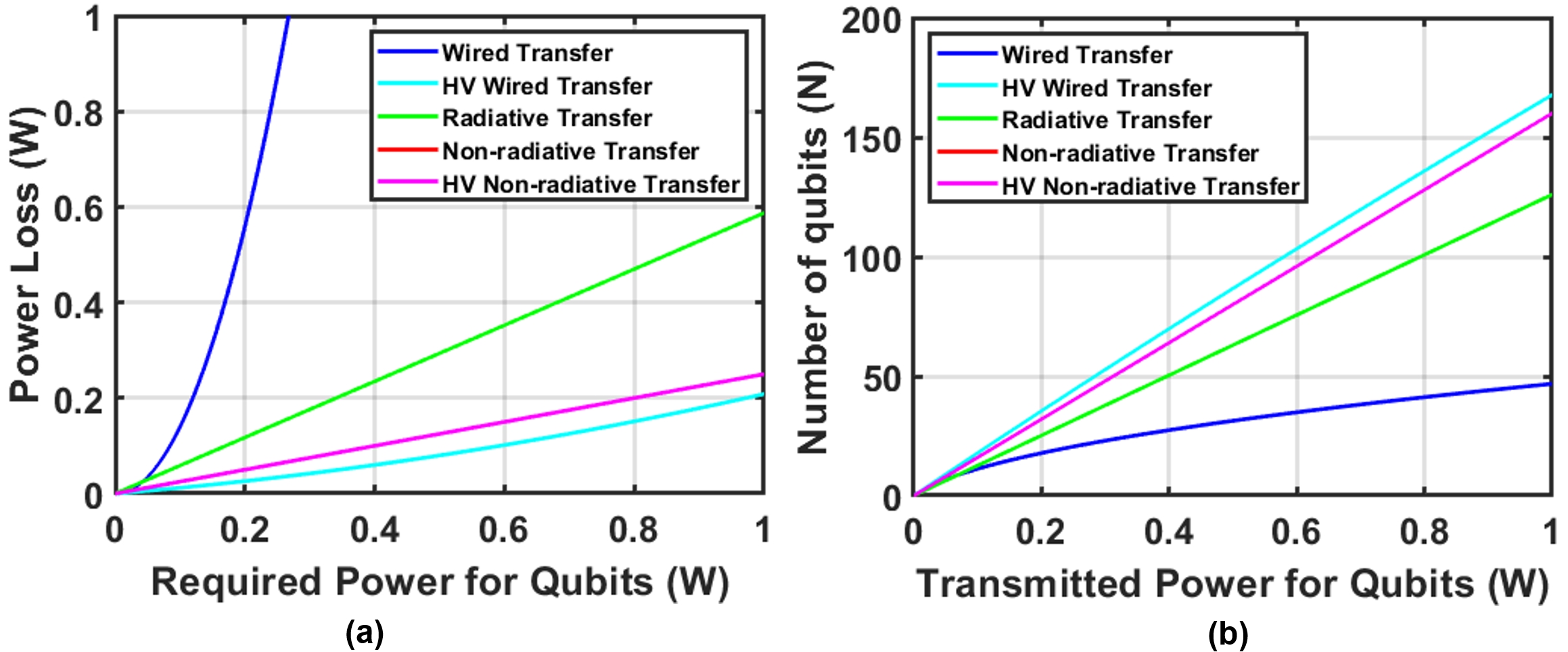}
\caption{Power comparison of different architectures: (a) power loss (b) the number of qubits that can be powered.}
\label{Power_comparison}
\end{figure}

Fig. \ref{Power_comparison} (a) compares the power loss of diffferent power transfer architectures (assuming power consumption 5mW per qubit). Conventional wired power transfer with one phosphor-bronze wire ($16\,\Omega @ 300\,\text{K},\ 12\,\Omega @ 4\,\text{K}$ per wire) exhibits nearly quadratic power loss, causing the power loss to become increasingly severe and even exceed the required delivered power if scaling. The efficiencies $ \eta_{\text{coup,ant}}$, $\eta_{\text{coup,coil}}$ and $ \eta_{\text{rad,r}}$ are assumed to be 70\%, 80\%, and 90\% respectively. Non-radiative and HV non-radiative transfer exhibit the same power loss when transmitter coil losses not contributing to heating are ignored. As illustrated in Fig.~\ref{Power_comparison}(b), when delivering the same total power of 1 W to qubits, the proposed four architectures can support approximately 3× more qubits than conventional wired transfer and wired transfer can achieve comparable loss performance only when employing 55 parallel wires.

\subsection{Heating leakage}
 The cooling system is designed to maintain the temperature $T_{\text{C}}$ lower than the ambient temperature $T_{\text{0}}$. The heating leakage in the dilution fridge would seriously worsen the burden of cooling power capacity\cite{martin2022}. The total heating leakage $Q_{\text{1}}$ at 4K arises from the thermal load of wires between warmer stage and cooler stage $P_{\text{load}}$, the worst-case transmission power loss entering the 4 K stage ($P_{\text{loss}}$), the heating leakage from ambient environment entering into the chamber $Q_{\text{0}}$, and the heating $Q_{\text{E}}$ caused by electronics at 4 K shown in Fig. \ref{Power_transfer_in_DF}. 
\begin{equation}
Q_{\text{1}} = P_{\text{load}}+P_{\text{loss}}+ Q_{\text{0}}+ Q_{\text{E}} \label{eq}
\end{equation}
The required cooling power $W_{\text{1}}$ can be expressed as:
\begin{equation}
W_{\text{1}} = \frac{Q_{\text{1}}}{COP(T_{\text{C}})}  = \frac{Q_{\text{1}}}{\eta_{\text{C}} \frac{T_{\text{C}}}{T_{\text{0}}-T_{\text{C}}} } \label{eq}
\end{equation}
where $COP(T_{\text{C}})$ is the Carnot coefficient of efficiency and $\eta_{C}$ is the correction factor of cryogenic cooling system.

The heating of the HV wired power transfer includes the thermal load between different stages $P_{\text{load}}$, the power loss caused by wires $P_{\text{loss, HV\_wired}}$ which is reduced compared to the conventional wired power transfer, heating leakage $Q_{\text{0}}$, and heating $Q_{\text{E}}$ caused by electronics including DC/DC converter, LDO, controller, etc. Compared to wired power transfer, additional heating of electronics is introduced by the DC/DC buck converter. For different topologies, the conversion efficiency $\eta_\text{DCDC}$ may vary; however, it can generally be approximated by accounting for conduction and switching losses as
\begin{equation}
\eta_\text{DCDC} \approx 
\frac{1}{
1 + 
\frac{I_{\text{out}}\left(R_{\text{HS}}D + R_{\text{LS}}(1-D) + R_{L}\right)}{V_{\text{out}}} 
+ 
\frac{0.5V_{\text{in}}(t_{r} + t_{f})f_{\text{sw}}}{V_{\text{out}}}
}
\label{eq:eta}
\end{equation}
where $I_{\text{out}}$ is the converter output current; $R_{\text{HS}}$ and $R_{\text{LS}}$ denote the on-resistances of the high-side and low-side MOSFETs, respectively; $R_{L}$ is the inductor’s DC resistance; $D$ is the duty cycle of the high-side MOSFET; $V_{\text{out}}$ and $V_{\text{in}}$ represent the output and input voltages; $t_{r}$ and $t_{f}$ are the rise and fall times of the high-side switching transitions; and $f_{\text{sw}}$ is the converter switching frequency. As the transmitted input voltage increases, $\eta_{\text{DCDC}}$ decreases, resulting in higher heat dissipation. Nevertheless, when the transmitted voltage is kept below 20 V, the loss can be restricted to under 10\%, which is considered negligible.

Radiative, non-radiative, and HV non-radiative transfer benefit from the absence of thermal load $P_{\text{load}}$ caused by wires. The heating in these architectures is primarily caused by transmission losses, denoted as $P_{\text{loss,rad}}$, $P_{\text{loss,non-rad}}$, and $P_{\text{loss,comb}}$, respectively. Consequently, the heating trend at 4 K follows the power loss, and HV non-radiative transfer outperforms HV wired transfer by eliminating thermal load.

\subsection{Noise}
Commercial power supplies at RT inherently generate frequency-dependent noise, dominated by $1/f$ noise at low frequencies and flattening to a white-noise floor at higher frequencies. Consequently, wired and HV wired power transfer exhibit higher DC noise, whereas radiative, non-radiative, and HV non-radiative architectures achieve a $10^{3}$–$10^{4}$ lower noise density at high frequencies (e.g., Keysight B296XA).

In HV wired power transfer, generating a higher supply voltage (e.g., 20 V) compared to conventional wired transfer (e.g., 2 V) results in approximately a 10× improvement in noise density. Additional switching noise introduced by DC/DC buck converters can further degrade performance, leading to inferior PSRR for cryogenic electronics.

Beyond the flicker-noise corner, the noise floor becomes flat and largely independent of the supply voltage. As a result, the noise of the rest transfer are determined mainly by the AC/DC or DC/DC conversion stages and are therefore similarly low.

\subsection{Power density and scalability}
Saving physical space in fridges directly benefits future scalability and accordingly improves power density. Wired and HV wired power transfer both exhibit low power density, as interconnect wiring occupies significant space and is difficult to re-route or re-drill when scaling to larger systems. HV wired transfer exhibits the highest power density, as the power loss in the transmitter coil is minimal.

\subsection{Reliability and complexity}
Wired power transfer is generally more reliable, as it provides stable power delivery. HV wired power transfer offers similar reliability but exhibits a higher aging probability due to HV device in the DC/DC converter in the long term, especially under repeated thermal cycling. In contrast, radiative, non-radiative, and HV non-radiative architectures require precise coil or antenna alignment, and the inclusion of additional components increases overall system complexity. All frequency ranges can cause electromagnetic interference to qubits, but $\sim$GHz noise is the direct threat since it resonates with qubit transitions, while lower and higher frequencies in non-radiative and radiative transfer mainly influence dephasing or heating, which can be suppressed by the shielding.

\section{Conclusions}
For large-scale power delivery quantum applications, several potential architectural approaches have been explored, taking into account thermal load, power loss, heating, coupling noise, power density, scalability, reliability, and complexity. HV wired power transfer delivers high voltage and low current from RT to electronics at CT, benefiting from reduced power loss ($I^{2}R_{\text{wire}}$) and lower system complexity. However, it still suffers from coupling noise, thermal load between stages and limited scalability due to wiring constraints. Both radiative and non-radiative approaches eliminate the thermal load constraint, wire-related losses and losses at the receiver antenna or coil, offer better heating performance, reduce coupling noise at high frequencies, and enable high power density without dense wiring. Non-radiative power transfer further reduces transmission loss at lower coupling frequencies compared to radiative power transfer. To further optimize heating, noise, and power density, a hybrid solution combining HV and non-radiative power transfer with superior heating performance presents a promising candidate for scalable quantum systems.

\section*{Acknowledgment}
All authors acknowledge support from the EU, grant 
HORIZON-EIC-2022-PATHFINDEROPEN-01-101099697, QUADRATURE.

\bibliographystyle{IEEEtran}
\bibliography{Bibliography.bib}

\end{document}